\documentclass[preprint*]{JHEP}
\usepackage[T1]{fontenc}
\usepackage{graphics}

\makeatletter

\providecommand{\LyX}{L\kern-.1667em\lower.25em\hbox{Y}\kern-.125emX\@}
\newcommand{\noun}[1]{\textsc{#1}}
\newcommand\fverb{\setbox\pippobox=\hbox\bgroup\verb}
\newcommand\fverbdo{\egroup\medskip\noindent%
			\fbox{\unhbox\pippobox}\ }
\newcommand\fverbit{\egroup\item[\fbox{\unhbox\pippobox}]}
\newbox\pippobox
\usepackage{latexsym}
\makeatother

\title{The neutron star in the Relativistic Mean-Field Theory }
\author{by Ryszard Ma\'{n}ka
\thanks{email:\email{manka@us.edu.pl}}, 
Ilona Bednarek\thanks{email:\email{bednarek@us.edu.pl}},  
Grzegorz Przyby\l {}a\\
University of Silesia, Institute of Physics, Katowice 40007,
ul. Uniwersytecka 4, Poland. }
\date{\today}
\abstract{The nucleon effective mass, binding energy and neutron star configuration in the Relativistic Mean-Field
Theory (RMF) is considered. The calculation are motivated by the construction
of the equation of state for a neutron star in the RMF. The equation of state
for the parameters set TM1 is calculated using the Feynman - Bogolubov variational
method for temperature different from zero. The structure of the neutron star is presented. The maximal stable
configuration are obtained for: 
$ M_{max}=1.91\, M_{\odot },\, \, \, R=12.84\, km.$ }

\preprint{AstroUSl-2000-1}

\begin{document}
\section*{Introduction}

The physics of compact objects like neutron stars offers an intriguing interplay
between nuclear processes and astrophysical observables. Neutron stars exhibit
conditions far from those encountered on earth. The determination of an equation
of state (EoS) for dense matter is essential to calculations of neutron star
properties. \\
This paper presents a basic model of neutron star \cite{hei} matter including
interactions among nucleons in the relativistic mean field approximation \cite{rei}\cite{walecka}\cite{weber}.
Especially the Walecka model (QHD) and its nonlinear extensions have been quite
successful and widely used for the description of hadronic matter and finite
nuclei. Increasing interest in neutron matter at finite temperature has been
observed recently in relation to the problems of hot neutron stars and of protoneutron
stars and their evolutions in particular. Theories concerning protoneutron stars
are being discussed in works by Prakash et. al. \cite{prakash}. Recently a
detail calculation has been done with different models to study the properties
of neutron stars \cite{prakash}. Glendenning \cite{glen} has studied the properties
of neutron star in the framework of nuclear relativistic field theory. In our
calculations, we used the TM1 \cite{tm1} parameter set, which has a capability
to reproduce the known results of finite nuclei as well as of normal nuclear
matter. The TM1 model possess a distinctively stiffer EOS.

\section*{The Relativistic Mean Field Theory}

The fields of the model RMF for \( \sigma ,\, \omega  \) and \( \rho  \)-mesons
are denoted as \( \varphi  \), \( \omega _{\mu } \), \( \rho _{\mu } \).
The Lagrange density function for this model has the following form 
\begin{eqnarray}
{\mathcal{L}} & = & \frac{1}{2}\partial _{\mu }\varphi \partial ^{\mu }\varphi -\frac{1}{4}R_{\mu \nu }^{a}R^{a\mu \nu }-\frac{1}{4}F_{\mu \nu }F^{\mu \nu }+\frac{1}{2}M^{2}_{\omega }\, \omega _{\mu }\omega ^{\mu }+\frac{1}{2}M^{2}_{\rho }\, \rho ^{a}_{\mu }\rho ^{a\mu }\label{lag} \\
 & - & U(\varphi )+\frac{1}{4}c_{3}(\omega _{\mu }\omega ^{\mu })^{2}+i\overline{\psi }\gamma ^{\mu }D_{\mu }\psi -\overline{\psi }(M-g_{s}\varphi )\psi +\nonumber \\
 &  & i\sum ^{2}_{f=1}\overline{L_{f}}\gamma ^{\mu }\partial _{\mu }L_{f}-\sum ^{2}_{f=1}g_{f}(\overline{L}_{f}He_{Rf}+h.c.)
\end{eqnarray}
 where 
\begin{equation}
R_{\mu \nu }^{a}=\partial _{\mu }\rho ^{a}_{\nu }-\partial _{\nu }\rho ^{a}_{\mu }+g\varepsilon _{abc}\rho _{\mu }^{b}\rho _{\nu }^{c}
\end{equation}
\begin{equation}
F_{\mu \nu }=\partial _{\mu }\omega _{\nu }-\partial _{\nu }\omega _{\mu }
\end{equation}
\begin{equation}
D_{\mu }=\partial _{\mu }+\frac{1}{2}ig_{\rho }\rho ^{a}_{\mu }\sigma ^{a}+ig_{\omega }\omega _{\mu }
\end{equation}
 The potential is given by 
\begin{equation}
U(\varphi )=\frac{1}{2}m^{2}_{s}\varphi ^{2}-\frac{1}{3}g_{2}\varphi ^{3}-\frac{1}{4}g_{3}\varphi ^{4}=\frac{1}{2}m^{2}_{s}\varphi ^{2}+\frac{1}{3!}\kappa \varphi ^{3}+\frac{1}{4!}\lambda \varphi ^{4}
\end{equation}
 The fermion fields are composed of protons, neutrons and electrons, muons and
neutrinos
\begin{equation}
\psi =\left( \begin{array}{l}
\psi _{p}\\
\psi _{n}
\end{array}\right) ,\, \, \, L_{1}=\left[ \begin{array}{l}
\nu _{e}\\
e^{-}
\end{array}\right] _{L},\, \, \, L_{2}=\left[ \begin{array}{l}
\nu _{\mu }\\
\mu ^{-}
\end{array}\right] _{L},\, \, \, e_{Rf}=(e^{-}_{R},\, \mu ^{-}_{R}).
\end{equation}
 \( M \) is the nucleon mass and \( m_{s} \), \( M_{\omega } \), \( M_{\rho } \)
are masses assigned to the mesons fields, \( g \), \( g' \) and \( g_{s} \)
are coupling constants. The Lagrangian function includes also the nonlinear
term \( \frac{1}{4}c_{3}(\omega _{\mu }\omega ^{\mu })^{2} \) which affects
remarkably the form of the equation of state. the Higgs field \( H \) takes
the form of 
\begin{equation}
H=\frac{1}{\sqrt{2}}\left( \begin{array}{l}
0\\
V
\end{array}\right) 
\end{equation}
 takes here the residual form.
\begin{table}
{\centering \begin{tabular}{|c|c|c|c|c|}
\hline 
Parameter&
L2\cite{walecka}&
NBL\cite{walecka}&
LN1\cite{toki}&
TM1\cite{tm1}\\
\hline 
\hline 
\( M \)&
\( 938\, MeV \)&
\( 938\, MeV \)&
\( 938\, MeV \)&
\( 938\, MeV \)\\
\hline 
\( M_{w} \)&
\( 786\, MeV \)&
\( 786\, MeV \)&
\( 795.359\, MeV \)&
\( 783\, MeV \)\\
\hline 
\( M_{\rho } \)&
\( 770\, MeV \) &
\( 770\, MeV \)&
\( 763\, MeV \)&
\( 770\, MeV \)\\
\hline 
\( m_{s} \)&
\( 500\, MeV \)&
\( 510\, MeV \)&
\( 492\, MeV \)&
\( 511.198\, MeV \)\\
\hline 
\( g_{2}=\kappa /2 \)&
\( 0 \)&
\( 2.03\, fm^{-1} \)&
\( 12.172\, fm^{-1} \)&
\( 7.2325\, fm^{-1} \)\\
\hline 
\( g_{3}=\lambda /6 \)&
\( 0 \)&
\( 1.666 \)&
\( -36.259 \)&
\( 0.6183 \)\\
\hline 
\( g_{s} \)&
\( 10.0773 \)&
\( 9.6959 \)&
\( 10.138 \)&
\( 10.0289 \)\\
\hline 
\( g_{\omega } \)&
\( 13.8655 \)&
\( 12.5889 \)&
\( 13.285 \)&
\( 12.6139 \)\\
\hline 
\( g_{\rho } \)&
\( 8.48784 \)&
\( 8.544 \)&
\( 4.6322 \)&
\( 4.6322 \)\\
\hline 
\( c_{3} \)&
\( 0 \)&
\( 0 \)&
\( 0 \)&
\( 71.3075 \)\\
\hline 
\end{tabular}\par}

\caption{\label{tab1}Parameters Set for the Lagrangian (\ref{lag}).}
\end{table}
The parameters used in NBL model \cite{walecka} are \( m_{s}=500 \) MeV with
\( \kappa =800 \) MeV and \( \lambda =10 \). 
\begin{figure}
{\par\centering \resizebox*{10cm}{!}{\includegraphics{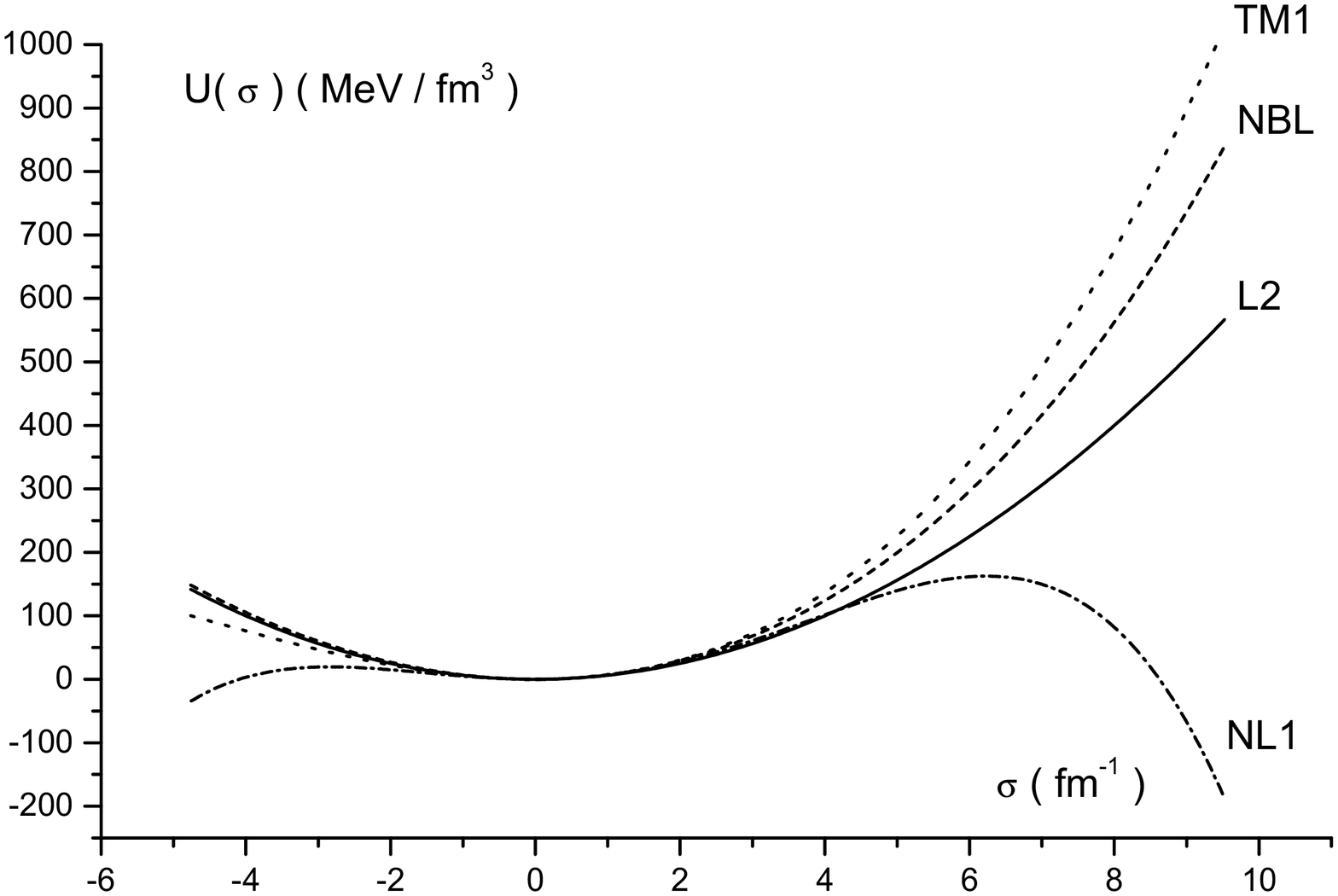}} \par}

\caption{\label{fig1}The potential \protect\( U(\varphi )\protect \) in the relativistic
mean field theory for L2 \cite{walecka}, NBL\cite{walecka}, LN1 \cite{toki},
TM1\cite{tm1} phenomenological parameters.}
\end{figure}
 The Euler equation for \( \Phi _{A}=\{ \)\( \varphi  \), \( \omega _{\mu } \),
\( \rho _{\mu } \),\( \psi \} \) fields are 
\begin{equation}
\label{egg1}
\Box \varphi =m_{s}^{2}\varphi +g_{2}\varphi ^{2}+g_{3}\varphi ^{3}-g_{s}\overline{\psi }\psi 
\end{equation}
 
\begin{equation}
\label{egg3}
-\partial _{\mu }F^{\mu \nu }=M^{2}_{\omega }\omega ^{\nu }+c_{3}(\omega _{\mu }\omega ^{\mu })\omega ^{\nu }-g_{\omega }J_{B}^{\nu }
\end{equation}
where 
\begin{equation}
J_{B}^{\nu }=\overline{\psi }\gamma ^{\nu }\psi 
\end{equation}
is the baryon current
\begin{equation}
\label{egg2}
-D_{\mu }R^{\mu \nu ,a}=M^{2}_{\rho }\, \rho ^{\nu ,a}-g_{\rho }J_{3}^{\nu }
\end{equation}
and 
\begin{equation}
J^{\nu }_{3}=\frac{1}{2}\overline{\psi }\gamma ^{\nu }\sigma ^{3}\psi 
\end{equation}
is the isospin current. In the system we have conservation of baryon charge
\[
Q_{B}=\int d^{3}xJ^{0},\]
and the isospin charge
\[
Q_{3}=\int d^{3}xJ^{0}_{3}\, .\]
The last is the Dirac equation 
\begin{equation}
\label{egg4}
i\gamma ^{\mu }D_{\mu }\psi -(M-g_{s}\varphi )\psi =0.
\end{equation}
The physical system is totally defined by the thermodynamic potential\cite{fet}
\begin{equation}
\Omega =-kTlnTr(e^{-\beta (H-\mu Q_{B}-\mu _{3}Q_{3})})
\end{equation}
 where H is the Hamiltonian of the physical system 
\begin{equation}
H=\sum _{A}\int d^{3}x\{\partial _{0}\Phi _{A}\pi ^{A}_{\Phi }-\mathcal{L}\}
\end{equation}
 and \( \pi ^{A}=\frac{\partial \mathcal{L}}{\partial (\partial _{0}\Phi _{A})} \)
is a momentum connected to the field \( \Phi _{A} \). The fields \( \Phi _{A}=\{ \)\( \varphi  \),
\( \omega _{\mu } \), \( \rho _{\mu } \),\( \psi \} \) denote all fields
in the system. The average charges 
\begin{equation}
\frac{\partial \Omega }{\partial \mu }=-<Q_{B}>,\, \, \, \, \frac{\partial \Omega }{\partial \mu _{3}}=-<Q_{3}>
\end{equation}
 can be obtained from the thermodynamic potential, of course they should be
conserved. In this paper we shall use the effective potential approach build
using the Bogolubov inequality \cite{rm}
\begin{equation}
\Omega \leq \Omega _{1}=\Omega _{0}(m_{B},m_{F})+<H-H_{0}>_{0}
\end{equation}
 \( \Omega _{0} \) is the thermodynamic potential of the trial system as effectively
free quasiparticle system described by the Lagrange function 
\begin{eqnarray}
{\mathcal{L}}_{0}(m_{B},m_{F}) & = & \frac{1}{2}\partial _{\mu }\overline{\varphi }\partial ^{\mu }\overline{\varphi }-\frac{1}{2}m_{B}^{2}\overline{\varphi }^{2}-\frac{1}{4}\overline{G}_{\mu \nu }^{a}\overline{G}^{a\mu \nu }-\frac{1}{4}\overline{F}_{\omega .\mu \nu }\overline{F}_{\omega }^{\mu \nu }\nonumber \\
 &  & +\frac{1}{2}M_{\omega }^{2}\overline{\omega }_{\mu }\overline{\omega }^{\mu }+\frac{1}{2}M_{\rho }^{2}\overline{\rho }_{\mu }^{a}\overline{\rho }^{a\mu }+\overline{\psi }(i\gamma ^{\mu }\overline{D}_{\mu }-m_{F})\psi 
\end{eqnarray}
 Similar to the general case 
\[
\overline{G}_{\mu \nu }^{a}=\partial _{\mu }\overline{\rho }^{a}_{\nu }-\partial _{\nu }\overline{\rho }^{a}_{\mu }\]
and 
\[
\overline{F}_{\omega ,\mu \nu }=\partial _{\mu }\overline{\omega }_{\nu }-\partial _{\nu }\overline{\omega }_{\mu }.\]
We decompose the \( \Phi _{A} \) field into two components, the effectively
free quasiparticle field \( \tilde{\Phi }_{A} \) and the classical boson condensate
\( \xi _{A} \)
\begin{equation}
\Phi _{A}=\tilde{\Phi }_{A}+\xi _{A}
\end{equation}
 In the case of the RMF model we have
\begin{equation}
\varphi =\overline{\varphi }+\sigma 
\end{equation}
\begin{equation}
\omega _{\mu }=\overline{\omega }_{\mu }+w_{\mu },\, \, w_{\mu }=\delta _{\mu ,0}w
\end{equation}
\begin{equation}
\rho _{\mu }^{a}=\overline{\rho }^{a}+r^{a}_{\mu },\, \, r^{a}_{\mu }=\delta ^{a,3}\delta _{\mu ,0}\, r
\end{equation}
 The \( \xi _{A}=\{\sigma ,\, w,\, r\} \) field will be treated as the variational
parameters in the effective potential. Also the boson and fermion mass \( m_{B},\, m_{F} \)
will be treated as as the variational parameters. The covariant derivative for
the trial system is 
\begin{equation}
\overline{D}_{\mu }=\partial _{\mu }+\frac{1}{2}ig_{\rho }r^{a}_{\mu }\sigma ^{a}+ig_{\omega }w_{\mu }
\end{equation}
This introduce the homogenous fermion interaction with boson condensate \( w_{\mu },\, r^{a}_{\mu }. \)
The fermion quasiparticle will obey the Dirac equation
\begin{equation}
(i\gamma ^{\mu }\, \overline{D}_{\mu }-m_{F})\psi =0
\end{equation}
 The constant condensate \( w,\, r \) simply shift the chemical potential from
\( \mu _{i}=\mu  \)\( ^{0}_{i} \) (when \( w=r=0 \)) to
\begin{eqnarray}
\mu _{n}=\mu ^{0}_{n}+\frac{1}{2}g_{\rho }r-g_{\omega }w &  & \label{mun} \\
\mu _{p}=\mu ^{0}_{p}-\frac{1}{2}g_{\rho }r-g_{\omega }w &  & \label{mup} 
\end{eqnarray}
where \( \mu _{n}=\mu +\frac{1}{2}\mu _{3} \) and \( \mu _{p}=\mu -\frac{1}{2}\mu _{3} \).\\
Neutrons, protons and electrons are in \( \beta  \)-equilibrium which can be
described as a relation among their chemical potentials 
\begin{equation}
\label{eq3}
\mu _{p}+\mu _{e}=\mu _{n}
\end{equation}
 where \( \mu _{p} \), \( \mu _{n} \) and \( \mu _{e} \) stand for proton,
neutron and electron chemical potentials respectively. If the electron Fermi
energy is high enough (greater then the muon mass) in the neutron star matter
muons start to appear as a result of the following reaction
\begin{equation}
e^{-}\rightarrow \mu ^{-}+\nu _{e}+\overline{\nu _{\mu }}
\end{equation}
 The chemical equilibrium between muons and electrons can be described by the
condition
\begin{equation}
\mu _{\mu }=\mu _{e}
\end{equation}
 When neutrinos are trapped inside the protoneutron star also the neutrino chemical
potential should be included 
\begin{equation}
\label{eq3a}
\mu _{p}+\mu _{e}=\mu _{n}+\mu _{\nu _{e}}.
\end{equation}
The density of the thermodynamic potential \( f_{1}=\Omega _{1}/V \) is equal
to 
\begin{eqnarray}
 & f_{1}(m_{B},m_{F},\sigma ,w,r)= & \label{free} \\
 & f_{0}(m_{B},m_{F})+\frac{1}{2}<\overline{\varphi }^{2}>_{0}(m_{s}^{2}-m_{B}^{2})+\frac{1}{8}\lambda <\overline{\varphi }^{2}>^{2}_{0}+ & \nonumber \\
 & \frac{1}{2}\kappa <\overline{\varphi }^{2}>_{0}\sigma +\frac{1}{2}(m_{s}^{2}+\frac{1}{2}\lambda <\overline{\varphi }^{2}>_{0})\sigma ^{2}+ & \nonumber \\
 & \frac{1}{3!}\kappa \sigma ^{3}+\frac{1}{4!}\lambda \sigma ^{4}+<\overline{\psi }\psi >_{0}(g\sigma -m_{F})+.., & \nonumber 
\end{eqnarray}

\begin{equation}
f_{0}=f_{B}+f_{F}
\end{equation}
 where \( f_{B} \) is the boson free energy and \( f_{F} \) the fermion one.
For boson field the free energy is 
\begin{equation}
f_{B}=\frac{k_{B}T}{(2\pi )^{3}}\int d^{3}pln(1-e^{-\beta \omega (p)})
\end{equation}
 with \( \omega (p)=\sqrt{{\textbf {p}}^{2}+m_{B}^{2}} \). For fermions we
have 4 degree of freedom, 2 for spin 1/2 and 2 for particle - antiparticle distinguishing.
For one fermion field the free energy is equal to 
\begin{equation}
f_{F}=-\sum _{i=\{n,p\}}\frac{2k_{B}T}{(2\pi )^{3}}\int d^{3}p\{ln(1+e^{-\beta (\epsilon (p)-\mu _{i})})+ln(1+e^{-\beta (\epsilon (p)+\mu _{i})})\}
\end{equation}
 now with \( \epsilon (p)=\sqrt{{\textbf {p}}^{2}+m_{F}^{2}} \). Variation
\[
\frac{\partial f_{1}}{\partial m^{2}_{B}}=0,\, \, \, \frac{\partial f_{1}}{\partial m_{F}}=0\]
with respect to the trial system \( L_{0} \) gives
\begin{equation}
\label{mb}
m^{2}_{B}=m^{2}_{s}+2g_{2}\sigma +3g_{3}(\sigma ^{2}+<\overline{\varphi }^{2}>_{0}),
\end{equation}
\begin{equation}
\label{mf}
m_{F}=M\delta =M-g_{s}\sigma .
\end{equation}
 In the local equilibrium inside the star the free energy reaches the minimum
at \( \sigma  \). 

\begin{figure}
{\par\centering \resizebox*{10cm}{!}{\includegraphics{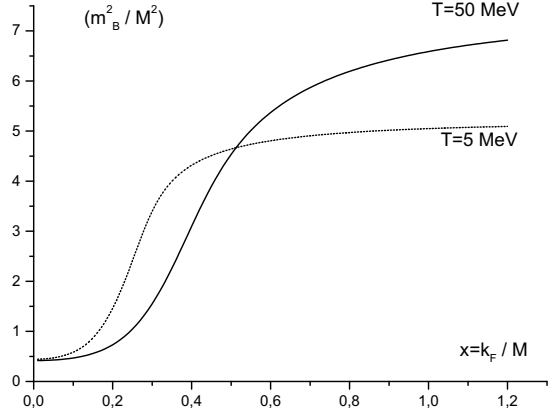}} \par}

\caption{\label{fig2}Meson mass \protect\( m^{2}_{B}/M^{2}\protect \) dependence on
the Fermi momentum \protect\( x=k_{F}/M\protect \) for temperature \protect\( T=50\, MeV\protect \).}
\end{figure}

The same result may be achieved calculating the averages of the equation of
motions (\ref{egg1}) for the effective system \( {\mathcal{L}}_{0} \). In
the mean field approximation the meson field operators are replaced by their
expectation values. We also consider the isotropic system at rest. For the scalar
field the equation (\ref{egg1}) have is follows
\begin{equation}
\label{eq1}
(m^{2}_{s}+\frac{1}{2}\lambda <\overline{\varphi }^{2}>_{0})\sigma +\frac{1}{2}\kappa \sigma ^{2}+\frac{1}{6}\lambda \sigma ^{3}=g_{s}<\overline{\psi }\psi >_{0}-\frac{1}{2}\kappa <\overline{\varphi }^{2}>_{0}
\end{equation}
 If we shall notice that 
\[
\frac{\partial f_{B}}{\partial m^{2}_{B}}=\frac{1}{2}<\overline{\varphi }^{2}>_{0},\]
 then it easy to obtain
\[
<\overline{\varphi }^{2}>_{0}=\frac{1}{2\pi ^{2}}\int \frac{dp\, p^{2}}{\sqrt{p^{2}+m_{B}^{2}}}\frac{1}{(exp(\beta \omega (p))-1)}\]
This result means that equation (\ref{mb}) is highly nonlinear with respect
to the meson mass \( m_{B} \) (Fig. \ref{fig2}). 
 As \( <\overline{\psi }\psi >_{0} \) (calculated with respect to \( {\mathcal{L}}_{0} \)
system) depends on the effective nucleon mass \( m_{F} \) (or \( \sigma  \)
) the equation (\ref{eq1}) is highly nonlinear also with respect to \( \sigma  \).
In most of papers \cite{7} temperature dependence of the boson fields is neglected.
Contrary to the fermion case the effective mass of boson \( m_{B} \) growing
with Fermi momentum. In the result the bosons temperature contributions may
be neglected.

Calculation similar to the boson case, base on the relation 
\[
\frac{\partial f_{F}}{\partial m_{F}}=<\overline{\psi }\psi >_{0}\]
gives
\begin{eqnarray}
<\overline{\psi }\psi >_{0}=\sum _{i=\{n,p\}}\frac{m_{F}}{\pi ^{2}}\int _{0}^{\infty }\frac{p^{2}dp}{\sqrt{p^{2}+m_{F}^{2}}}\{\frac{1}{\exp (\beta (\epsilon _{p}-\mu _{i}))+1}+ &  & \\
\frac{1}{\exp (\beta (\epsilon _{p}+\mu _{i}))+1}\}. &  & \nonumber 
\end{eqnarray}
The quantum average \( <\overline{\psi }\psi >_{0} \) depends on the neutron
and proton chemical potentials \( \mu _{p} \), \( \mu _{n} \). In the result
the effective nucleon effective mass \( m_{F} \) also will be dependent he
neutron and proton chemical potentials. In the result the solution \( \sigma  \)
of the equation (\ref{egg1}) also will be dependent on \( \xi  \). The same
situation will consider other fields. This model is the simple example of the
relativistic mean filed theory \cite{walecka}. 

\begin{figure}
{\par\centering \resizebox*{10cm}{!}{\includegraphics{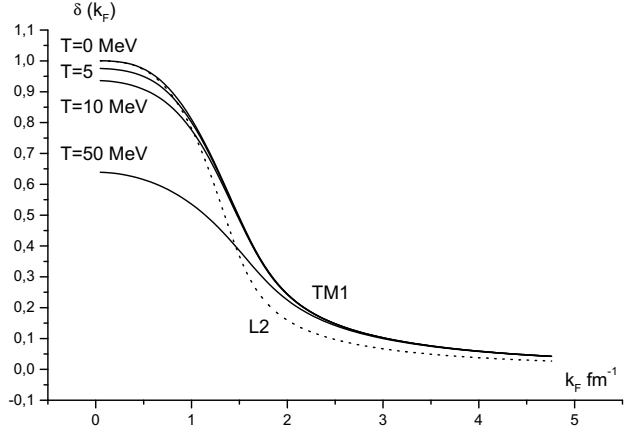}} \par}

\caption{\label{figd}The effective neutron mass \protect\( m_{F}=\delta (x_{F})M\protect \)
as function of the neutron Fermi momentum \protect\( x=k_{F}\, (\, fm^{-1})\protect \). }
\end{figure}

The effective mass \( m_{F} \) (or \( \delta =m_{F}/M \) ) dependence on the
dimensionless Fermi momentum \( x_{F} \) is presented on the Fig.\ref{figd}.
The binding energy 
\begin{equation}
E_{0}=\epsilon (x_{F},T)/Q_{B}-M
\end{equation}
for nucleon symmetric phase is presented on Fig. 2 (see Table \ref{tab2}).
\begin{table}
{\centering \begin{tabular}{|c|c|c|c|c|}
\hline 
Parameter&
L2\cite{walecka}&
NBL\cite{walecka}&
LN1\cite{toki}&
TM1\cite{tm1}\\
\hline 
\hline 
\( E_{0}\, MeV \)&
\( -15.74 \)&
\( -16.59 \)&
\( -16.42 \)&
\( -16.26 \)\\
\hline 
\( k_{F,0}\, fm^{-1} \)&
\( 1.30 \)&
\( 1.31 \)&
\( 1.309 \)&
\( 1.29 \)\\
\hline 
\( \rho _{0}\, fm^{-3} \)&
\( 0.149 \)&
\( 0.1524 \)&
\( 0.1517 \)&
\( 0.1452 \)\\
\hline 
\( \delta _{0}=m_{F}/M \)&
\( 0.54 \)&
\( 0.60 \)&
\( 0.573 \)&
\( 0.634 \)\\
\hline 
\noun{\( K\, MeV \)}&
\( 545.93 \)&
\( 445.73 \)&
\( 215.19 \)&
\( 281.53 \)\\
\hline 
\end{tabular}\par}

\caption{\label{tab2} The nucleon symmetric phase properties at the saturation point.}
\end{table}
We see that binding energy strongly depends on temperature and above \( T>15 \)
MeV is positive.
\begin{figure}
{\par\centering \resizebox*{10cm}{!}{\includegraphics{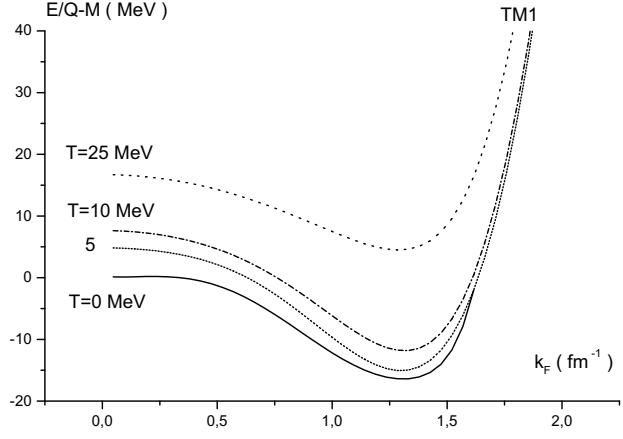}} \par}

\caption{\label{eot} The temperature dependence of the binding energy for nucleon symmetric
TM1 phase.}
\end{figure}
The temperature dependence of the binding energy in presented on the Fig \ref{eot}.
To calculate the properties of the neutron star we need the energy-momentum
tensor. In case of the fermions field it is more convenient to use the reper
filed \( e_{\mu }^{a} \) defined as follows \( g_{\mu \nu }=e_{\mu }^{a}e_{\nu }^{b}\eta _{ab} \)
where \( \eta _{ab} \) is the flat Minkowski space-time matrix. The general
definition 
\begin{equation}
T_{\mu \nu }=2\frac{\partial L_{B}}{\partial g^{\mu \nu }}+e^{a}_{\mu }\frac{\partial L_{F}}{\partial e^{a\nu }}-g_{\mu \nu }L
\end{equation}
allows us to calculate the density of energy and pressure. The total Lagrange
function \( L=L_{B}+L_{F} \) is divided into boson and fermion part. To calculate
the density of energy and pressure we shall average the energy-momentum tensor
\( T_{\mu \nu } \) with respect to the quasi equilibrium configuration defined
by the trial system \( L_{0} \). We define the density of energy and pressure
by the energy - momentum tensor
\begin{equation}
<T_{\mu \nu }>=(P+\epsilon )u_{\mu }u_{\nu }-Pg_{\mu \nu }
\end{equation}
where \( u_{\mu } \) is a unite vector (\( u_{\mu }u^{\mu }=1 \)). So, the
calculations give
\begin{equation}
\epsilon (x_{F},T)=\rho c^{2}=-\frac{1}{2}M^{2}_{\omega }w^{2}-\frac{1}{2}M^{2}_{\rho }r^{2}+g_{\omega }Q_{B}w+\frac{1}{2}g_{\rho }Q_{3}r-\frac{1}{4}c_{3}w^{4}+U(\sigma )+\epsilon _{F}
\end{equation}
\begin{equation}
P(x_{F},T)=\frac{1}{2}M^{2}_{\omega }w^{2}+\frac{1}{2}M^{2}_{\rho }r^{2}+\frac{1}{4}c_{3}w^{4}-U(\sigma )+P_{F}
\end{equation}
where
\begin{equation}
\epsilon _{F}=\epsilon _{0}\chi (x_{F},T)
\end{equation}
\begin{equation}
P_{F}=P_{0}\phi (x_{F},T)
\end{equation}
The fact that neutron mass depends on fermion concentration ( or neutron chemical
potential \( \mu  \)) now must be included into \( \chi (x_{F},T) \) and \( \phi (x_{F},T) \),
\\
\begin{eqnarray}
\chi (x_{F},T)=\frac{1}{^{\pi ^{2}}}\int _{0}^{\infty }dz\, z^{2}\sqrt{z^{2}+\delta ^{2}(x_{F})}\{\frac{1}{\exp ((\sqrt{\delta ^{2}(x_{F})+z^{2}}-\mu ')/\tau )+1} &  & \\
+\frac{1}{\exp ((\sqrt{\delta ^{2}(x_{F})+z^{2}}+\mu ')/\tau )+1}, &  & \nonumber 
\end{eqnarray}
\begin{eqnarray}
\phi (x_{F},T)=\frac{1}{3\pi ^{2}}\int _{0}^{\infty }\frac{z^{4}dz}{\sqrt{z^{2}+\delta ^{2}(x_{F})}}\{\frac{1}{\exp ((\sqrt{\delta ^{2}(x_{F})+z^{2}}-\mu ')/\tau )+1} &  & \\
+\frac{1}{\exp ((\sqrt{\delta ^{2}(x_{F})+z^{2}}+\mu ')/\tau )+1}\} &  & \nonumber 
\end{eqnarray}
where \( \tau =(k_{B}T)/M \), 
\begin{equation}
\label{mu}
\mu '=\mu /M=\sqrt{\delta ^{2}(x_{F})+x_{F}^{2}}
\end{equation}
 and 
\begin{equation}
\label{mui}
x_{F}=k_{F}/M
\end{equation}
Similar to paper \cite{toki} we have introduced (\ref{mu},\ref{mui}) the
dimensionless ``Fermi'' momentum even at finite temperature which exactly
corresponds to the Fermi momentum at zero temperature. Both \( \epsilon _{F} \)
and \( P_{F} \) depend on the neutron chemical potential \( \mu  \) or Fermi
momentum \( x_{F} \). This parametric dependence on \( \mu  \) (or \( x_{F} \))
defines the equation of state. The various equations of state for different
parameters sets is presented on fig.\ref{feso}. 
\begin{figure}
{\par\centering \resizebox*{10cm}{!}{\includegraphics{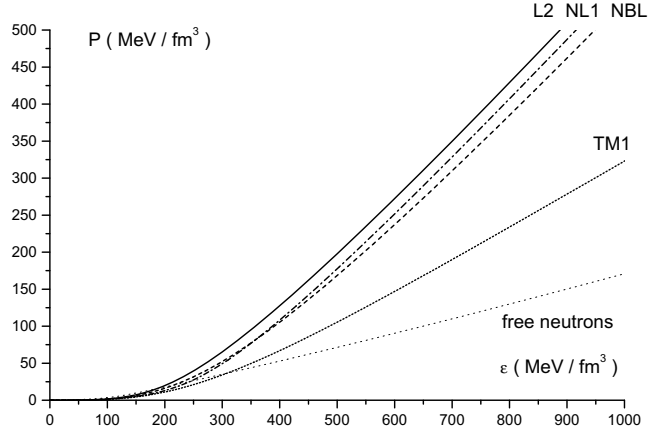}} \par}

\caption{\label{feso} The equation of 
state for different parameters sets (L2, NBL,TM1).}
\end{figure}
The equation of state for the parameters set TM1 for temperature T=10 MeV is
presented on fig.\ref{eost}. It is interesting to notice that even for \( Q_{B}=0 \)
due to the presence of the thermal exited particle antiparticle pairs there
is finite energy and pressure density.
\begin{figure}
{\par\centering \resizebox*{10cm}{!}{\includegraphics{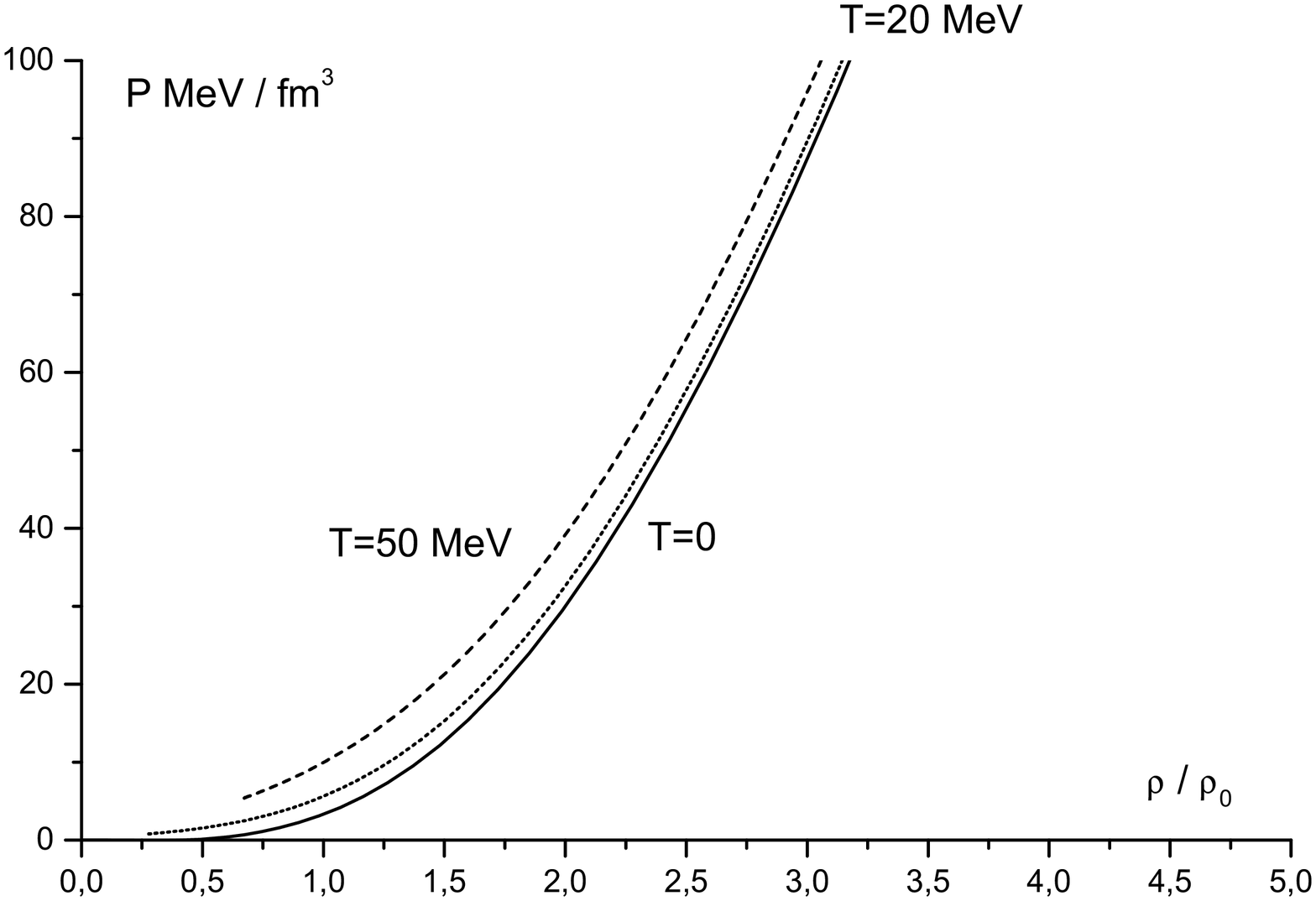}} \par}

\caption{\label{eost} The equation of state for the parameters 
set TM1 for T=0 (solid line) and T=20, 50 MeV temperature.}
\end{figure}
 It is interesting to notice that the meson field \( \varphi  \) as the scalar
field contributes to the pressure with negative sign while the vector meson
fields (\( \omega ,\, \rho  \)) with the positive one.

\section*{The neutron star}

\begin{figure}
{\par\centering \resizebox*{10cm}{!}{\includegraphics{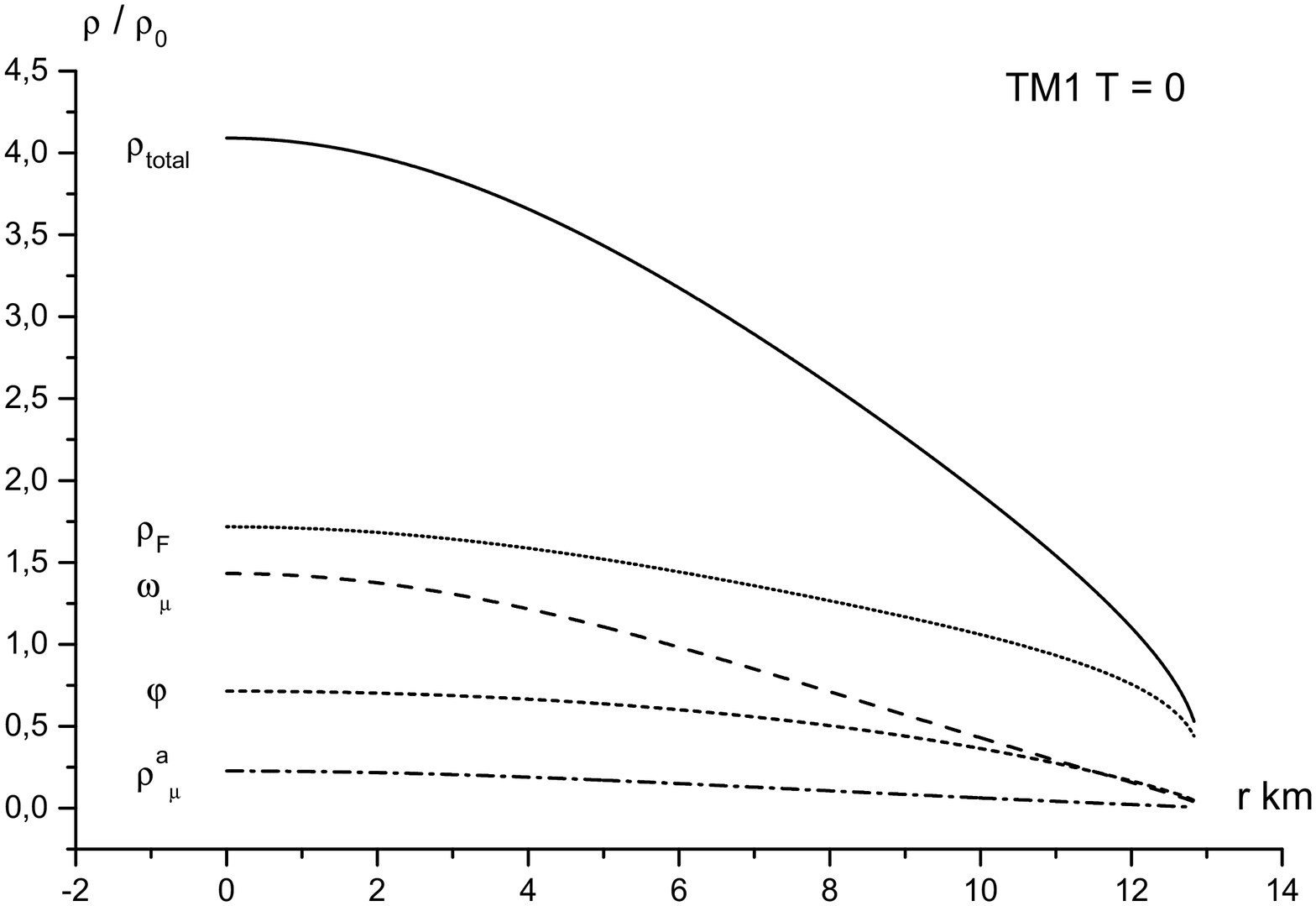}} \par}

\caption{\label{rfig1}The neutron star density profile for fermions \protect\( \rho _{F}\protect \)
scalar bosons \protect\( \varphi \protect \) and gauge bosons (\protect\( \omega _{\mu }\protect \),\protect\( \rho \protect \)\protect\( ^{a}_{\mu }\protect \)).}
\end{figure}

In this paper we present numerical results describing the structure of neutron
star based on the relativistic mean field theory. It is possible to describe
a static spherical star solving the OTV equation. 
\begin{equation}
\label{teq1}
\frac{dP(r)}{dr}=-\frac{G}{r^{2}}(\rho (r)+\frac{P(r)}{c^{2}})\frac{(m(r)+\frac{4\pi }{c^{2}}P(r)r^{3})}{(1-\frac{2Gm(r)}{c^{2}r})}
\end{equation}
\begin{equation}
\label{teq2}
\frac{dm(r)}{dr}=4\pi r^{2}\rho (r)
\end{equation}
 Having solved the OTV equation the pressure \( p(r) \), mass \( m(r) \) and
density \( \rho (r) \) were obtained. To obtain the total radius \( R \) of
the star the fulfillment of the condition \( p(R)=0 \) is necessary. This allows
to determine the total gravitational mass of the star \( M(R) \).\\
 Introducing the dimensionless variable \( \xi , \) which is connected with
the star radius \( r \) by the relation \( r=a\xi  \) enables to define the
functions \( p(r) \), \( \rho (r) \) and \( m(r) \)
\begin{equation}
\rho (r)=\rho _{0}\chi (x(\xi ))
\end{equation}
\begin{equation}
P(r)=P_{0}\varphi (x(\xi ))
\end{equation}
 
\begin{equation}
m(r)=M_{\odot }v(\xi )
\end{equation}
by \( \xi  \). If we define dimensionless functions.
\begin{equation}
\lambda =\frac{GM_{\odot }\rho _{c}}{P_{0}a},\, \, \mu =3\frac{M_{c}}{M_{\odot }},\, \, M_{c}=\frac{4}{3}\pi \rho _{0}a^{3}
\end{equation}
 are also need to achieve the OTV equation of the following form 
\begin{equation}
\label{volk1}
\frac{d\varphi }{d\xi }=-\lambda (\chi (\xi )+\varphi (\xi ))\frac{v(\xi )+\mu \varphi (\xi )\xi ^{3}}{\xi ^{2}(1-\frac{r_{g}}{a}\frac{v(\xi )}{\xi })}
\end{equation}
\begin{equation}
\label{volk2}
\frac{dv}{d\xi }=\mu \chi (\xi )\xi ^{2}
\end{equation}
 with \( r_{g} \) being the gravitational radius.The equations (\ref{volk1},\ref{volk2})
are easy integrated numerically, For example, for the neutron star with the
central density \( \rho _{c}=9\, 10^{14}\, g/cm^{3} \) the star profile in
the mean field approach is presented on the Fig.\ref{rfig1}.
It's interesting that fermions contribution to density is lower then a half
of total density. The biggest contributions come from the nucleons, the gauge
boson field \( \omega _{\mu } \) and scalar boson field \( \varphi  \).
\begin{figure}
{\par\centering \resizebox*{10cm}{!}{\includegraphics{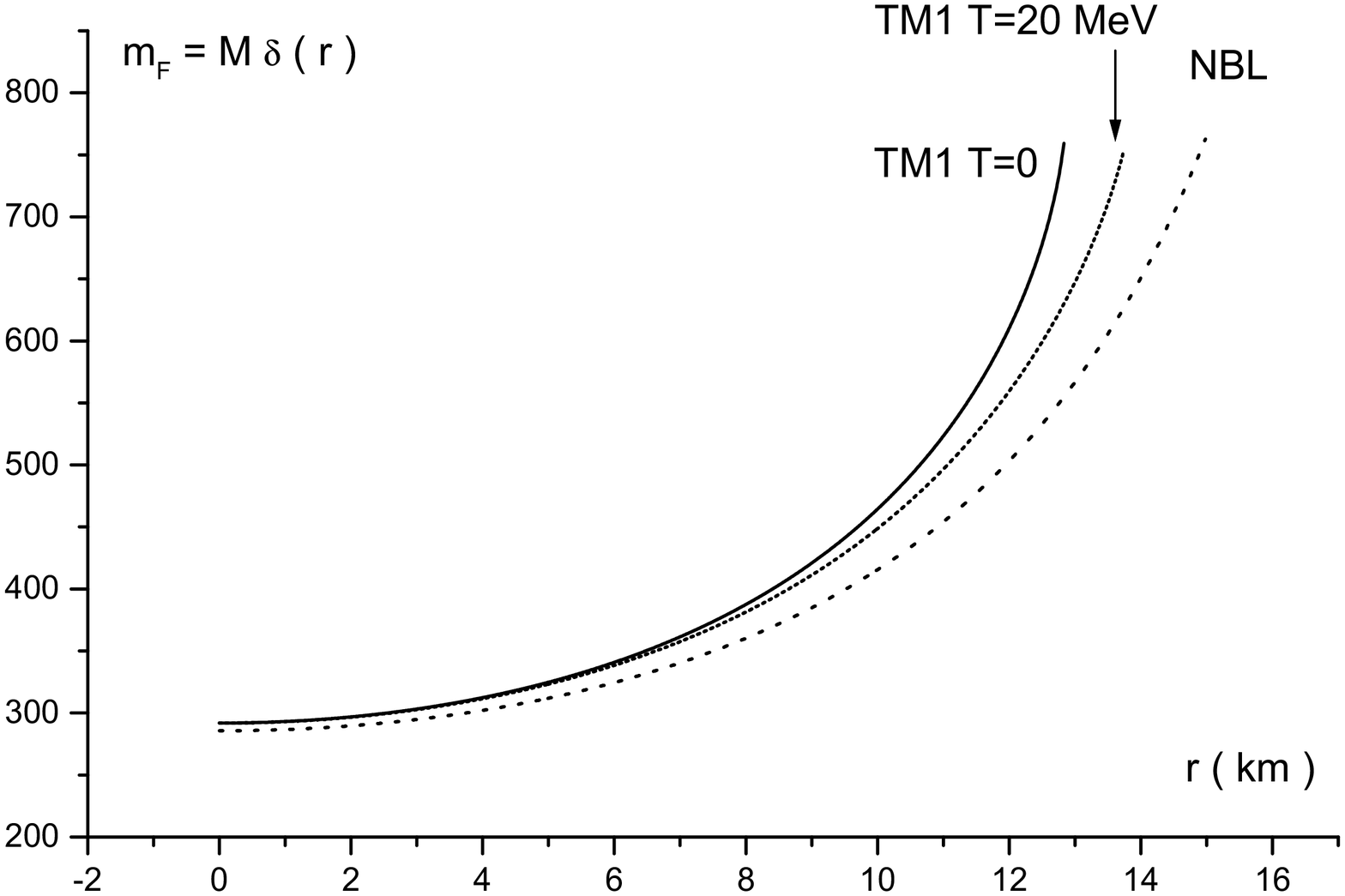}} \par}

\caption{\label{last}The effective neutron mass profile (\protect\( m_{N}=M\delta \protect \))
inside the star. }
\end{figure}
 On the surface of star we have non zero value of density (\( \rho \sim 0.25<\rho _{0} \))
and nucleon mass lower than vacuum nucleon mass (\( m_{F}\sim \, 762\, MeV<M \)).
Inside star with \( \rho _{c}=9\times 10^{14}\, g/cm^{3} \) (maximal stable
configuration for the TM1 parameters set) value of nucleon mass varies from
\( \approx 300 \) MeV in centrum of star to \( \approx 762 \) MeV on the surface
(Fig. \ref{last}). Fig. \ref{rhor}. shows the stellar masses as a function of the central
density. The parameters of the maximum mass configuration are: 
\begin{equation}
M_{max}=1.91\, M_{\odot },\, \, \, R=12.84\, km.
\end{equation}
 This fact is easy to notice on the mass-radius diagram (Fig. \ref{rmd}). 
\begin{figure}
{\par\centering \resizebox*{10cm}{!}{\includegraphics{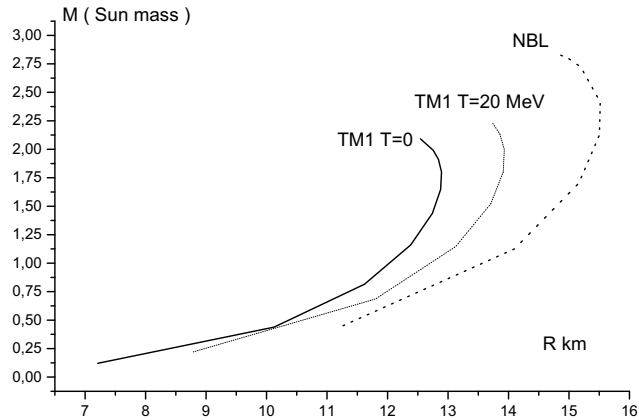}} \par}

\caption{\label{rmd}The R-M diagram for the neutron star. }
\end{figure}
When temperature is different from zero, a star is in generar bigger and more
massive (see Figs. \ref{rmd} and \ref{rhor}).
\begin{figure}
{\par\centering \resizebox*{10cm}{!}{\includegraphics{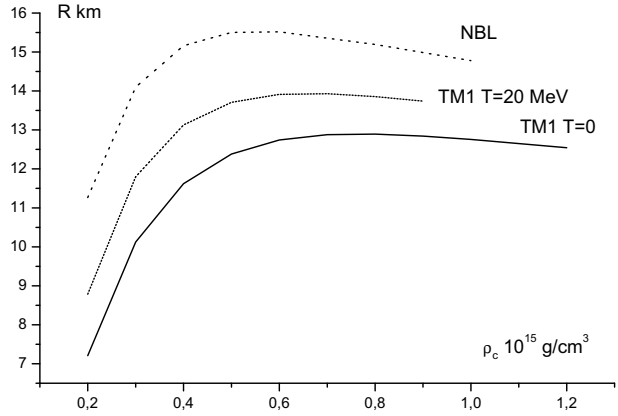}} \par}

\caption{\label{rhor}The neutron star radius dependence from the central density \protect\( \rho _{c}\protect \).}
\end{figure}

\section*{Conclusion}

The structure of static neutron stars can be determined by solving the Tolman-Oppenheimer-Volkoff
equations. The equation of state will strong influence on the neutron star properties.
The aim of this work has been to present the equation of state in the RMF for
the nuclear matter with temperature different from zero. The hot neutron star
in the simple L2 parameters set was examined in our previous work \cite{ilona}.
The main objective of our work was to study the influence of the temperature
on the main parameters of a neutron star. Neutron matter at finite temperature
is of increasing interest in relation to the problems of hot neutron stars.
In order to achieve the proper form of the equation of state the relativistic
mean field approach was involved. 

\newpage

\section*{Appendix: Nuclear matter properties}

\begin{figure}
{\par\centering \resizebox*{10cm}{!}{\includegraphics{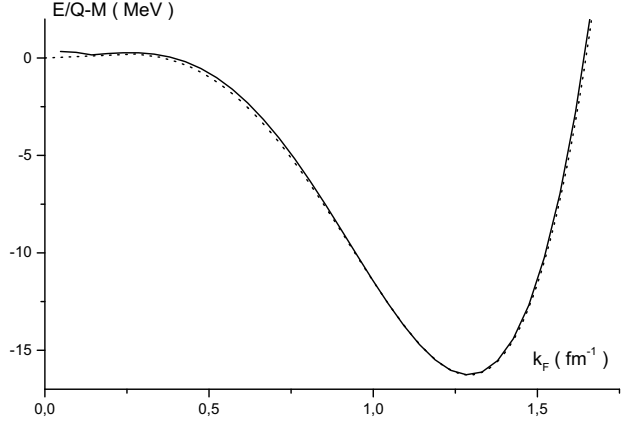}} \par}

\caption{\label{figeop}The binding energy for nucleon symmetric TM1 phase according
to the RMF approach (the solid line) and (\ref{eop}) (the dot line), respectively.}
\end{figure}

\begin{figure}
{\par\centering \resizebox*{10cm}{!}{\includegraphics{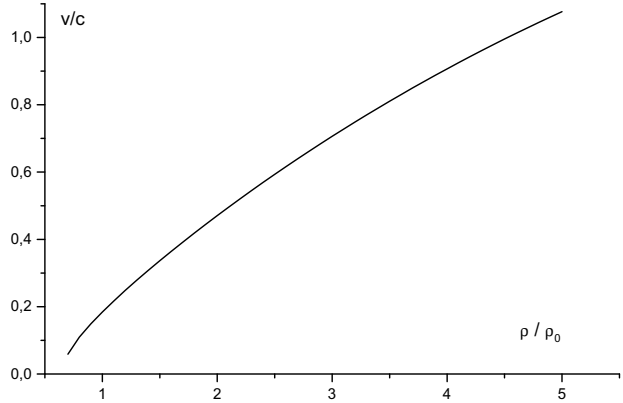}} \par}

\caption{\label{figv}The adiabatic sound speed as the function of the relative density
\protect\( u=\rho /\rho _{0}\protect \) for the TM1 parameters set.}
\end{figure}

The nuclear symmetric matter may be described by the phenomenological equation
of state \cite{pras}
\begin{equation}
\label{epsp}
\varepsilon (u)=\rho _{0}u(M+\frac{3}{5}\frac{\hbar ^{2}k_{F}^{2}}{2M}+\frac{1}{2}Au^{2}+\frac{B}{(\sigma +1)}u^{\sigma })
\end{equation}
where \( u=\rho /\rho _{0} \) is a dimensionless density. For symmetric nuclear
matter dimensionless ~\( \rho _{0}=2.5\, 10^{14}\,  \) \( g\, cm^{-3} \) \( =0.15\, nucleons/fm^{3} \)
\( =140 \)~ MeV\( \, fm^{-3} \). The parameter of \( \varepsilon  \) (\ref{epsp})
are 
\begin{equation}
\sigma =\frac{K_{0}+2E_{F,0}}{3E_{F,0}-9E_{0}},
\end{equation}

\begin{equation}
B=(\frac{\sigma +1}{\sigma -1})[\frac{1}{3}E_{F,0}-E_{0}],
\end{equation}

\begin{equation}
A=E_{0}-\frac{5}{3}E_{F,0}-B.
\end{equation}
\( E_{F,0} \) is the nonrelativistic nucleon Fermi energy at the saturation
point \( k_{F,0} \) (see Table \ref{tab2}). For the TM1 parameter set (see
Table \ref{tab1}) we have
\[
\sigma =1.5485;\, \, A=-1.58.52\, MeV;\, \, B=107.689\, MeV;\]
In this approach the binding energy (see Fig.\ref{figeop} ) is 
\begin{equation}
\label{eop}
E_{0}(u)=E_{F,0}u^{\frac{2}{3}}+\frac{1}{2}Au+\frac{B}{(\sigma +1)}u^{\sigma }.
\end{equation}
The pressure is 
\begin{equation}
P=u^{2}\frac{d(\varepsilon /Q_{B})}{d\rho }|_{\rho _{0}}=\frac{2}{3}E_{F,0}u^{\frac{5}{3}}+\frac{1}{2}Au^{2}+\frac{B\sigma }{(\sigma +1)}u^{\sigma +1}.
\end{equation}
 The minimum of the binding energy determine the equilibrium Fermi momentum
\( k_{F,0} \), density \( \rho _{0} \) and incompressibility factor 
\begin{equation}
K=9\rho _{0}\frac{d^{2}(\varepsilon /Q_{B})}{d\rho ^{2}}|_{\rho _{0}}.
\end{equation}
The empirical value of \( K \) is \( 210\pm 30\, MeV \) \cite{K}. The incompressibility
factor is equal to
\begin{equation}
K(u)=9\frac{d(P/\rho )}{du}|_{\rho _{0}}=10E_{F,0}u^{\frac{2}{3}}+9(Au+B\sigma u^{\sigma }).
\end{equation}
 The chemical potential defined for particle species \( i \) is given by 
\begin{equation}
\mu _{i}(u)=\frac{d(\varepsilon )}{d\rho _{i}.}.
\end{equation}
For example, for nucleon symmetric saturation point \( \mu =912.73\, MeV \)
for nucleons. The adiabatic sound speed
\begin{equation}
(\frac{v}{c})^{2}=\frac{dP}{d\varepsilon }=\frac{K}{9\mu }
\end{equation}
is presented on the Fig.\ref{figv}. At the saturation point we have the sound
speed \( v=0.184c \). For TM1 parameters, for \( \rho  \) above \( 5\rho _{0} \)
the theory looses its sense. For other parameters sets when \( K \) is grater
the range of the theory validity is smaller.
\sl{Acknowledgment}\\
 The authors are thankful to F. Weber  for helpful discussions.

\end{document}